\documentstyle[12pt,epsfig ]{article}
\newcommand {\eqref} [1] {(\ref {#1})}
\newcommand {\slsh} [1] {\not{\hbox{\kern-3pt${#1}$}}} 

\begin{document}
\newcommand{\beq}{\begin{equation}}
\newcommand{\eeq}{\end{equation}}
\def\beqn{\begin{eqnarray}}
\def\eeqn{\end{eqnarray}}

\newcommand{\Tr}{{\rm Tr}\,}
\newcommand{\E}{{\cal E}}

\newcommand{\ntwo}{${\cal N}=2\;$}
\newcommand{\none}{${\cal N}=1\;$}
\newcommand{\noneh}{${\cal N}=\,
^{\mbox{\small 1}}\!/\mbox{\small 2}\;$}
\newcommand{\vp}{\varphi}
\newcommand{\ve}{\varepsilon}
\newcommand{\pt}{\partial}

  \newcommand {\ber}{\begin{eqnarray*}}
  \newcommand {\eer} {\end{eqnarray*}}

\newcommand{\Nfour} {${\mathcal N}=4\ $}
\newcommand{\Ntwo}{${\mathcal N}=2\ $}
\newcommand{\None}{${\mathcal N}=1\ $}
\newcommand{\Ztwo}{${\bf Z}_2\ $}
\newcommand{\Zk}{${\bf Z}_k\ $}
\newcommand{\Dslash}{\,{\raise.15ex\hbox{/}\mkern-12mu D}}

\newcommand{\gsim}{\lower.7ex\hbox{$
\;\stackrel{\textstyle>}{\sim}\;$}}
\newcommand{\lsim}{\lower.7ex\hbox{$
\;\stackrel{\textstyle<}{\sim}\;$}}

\begin{flushright}{ SLAC-PUB-13516, FTPI-MINN-09/05, UMN-TH-2735/09}
\end{flushright}
\vskip 1.5cm

\begin{center}
{ {\large\bf Multiflavor QCD$^*$ on \boldmath{$R_3\times S_1$}: Studying Transition} \\[1mm]
{\large\bf  From Abelian to Non-Abelian Confinement}
}
 
 \end{center}
 
 \vspace{3mm}
\centerline{\large  M. Shifman${}^{a}$  and M. \"{U}nsal ${}^{b,c}$}

\vskip 0.2cm

\centerline{${}^a$   \footnotesize\it William I. Fine Theoretical Physics Institute,}
\centerline{\footnotesize\it University of Minnesota, Minneapolis, MN 55455}

\vskip 0.2cm

\centerline{${}^b$ \footnotesize\it SLAC, Stanford 
University, Menlo Park, CA 94025}
\vskip 0.2cm
\centerline{${}^c$ \footnotesize\it  Physics Department, 
Stanford University, Stanford, CA 94305 }

\vskip 1cm

\begin{abstract}
The center-stabilized multiflavor QCD* theories formulated on  $R_3\times S_1$
exhibit both Abelian and non-Abelian confinement 
as a function of the $S_1$ radius, similar to 
the Seiberg--Witten theory as a function of the mass deformation parameter.  
For sufficiently small number of flavors and small $r(S_1)$, 
we show occurence of a mass gap in gauge fluctuations, and linear confinement.  This is a regime  of  confinement  without continuous chiral  symmetry breaking  ($\chi$SB). 
 Unlike one-flavor theories where
  there is no phase transition in $r(S_1)$,  the multiflavor theories possess a single  
 phase  transition 
  associated   with breaking of the continuous $\chi$S.
We conjecture that the scale of the $\chi$SB  is parametrically tied 
up with the scale of Abelian to non-Abelian confinement transition. 

\end{abstract}

\newpage

\section{Introduction}

In supersymmetric \ntwo Yang--Mills theories slightly deformed to \none by a
mass term $\mu\Tr\Phi^2$  for chiral superfield  $\Phi$ linear confinement is a result of 
the dual Meissner effect \cite{SW1}.  
In the limit of small $\mu$ amenable to analytic studies
\cite{SW1} confinement is  Abelian (for a definition of Abelian vs. non-Abelian 
confinement see e.g. \cite{SU1}). Many theorists believe that in passing to
large  $\mu$, (i.e. $\mu\gsim \Lambda$), a smooth transition to non-Abelian confinement 
-- pertinent to pure \none supersymmetric Yang-Mills -- takes place in the Seiberg--Witten model. This is sometimes referred to as
the same universality class hypothesis.
In non-supersymmetric theories a  construction serving the same purpose
-- studying the transition from  Abelian to non-Abelian confinement by tuning 
 an adjustable parameter -- was engineered in \cite{SU2} (see also \cite{SU3}). There we considered SU$(N)$ Yang--Mills theories on 
$R_3\times S_1$ treating the radius of the compact dimension $r(S_1)$
as a free parameter. At small  $r(S_1)$ (i.e. $r(S_1)\ll \Lambda^{-1}$)
 we introduced a double-trace deformation
stabilizing the vacuum of the theory at a center-symmetric point.
With this stabilization,  the Polyakov mechanism \cite{Pol} guarantees linear Abelian confinement both, in pure Yang--Mills theories
and in those with one massless quark in various representations of the SU$(N)$
gauge group \cite{SU2}. A discrete chiral symmetry ($\chi $S) inherent to two-index representations is  spontaneously broken. Both effects, linear
confinement and $\chi $SB were caused by topological excitations
in the vacuum (monpole-instantons, bions, etc.) which are  under complete control at small  $r(S_1)$. No obvious phase  transitions in passing from the weak coupling Abelian regime at small-$r(S_1)$  to the strong coupling
decompactification/non-Abelian confinement regime at  $r(S_1)\gg \Lambda^{-1}$
was detected. 
The trace of the Polyakov line ${\mathcal U} = P\,exp\,\left\{i
\int_0^L\, A_z\,dz\right\}$
remains vanishing in both regimes.
Thus, the same universality class hypothesis is not necessarily tied up with supersymmetry.

In this paper we turn to SU$(N)$ gauge theories with several flavors,
within the same theoretical framework.\footnote{Neither the number of colors $N$
nor the number of flavors $N_f$ are asumed to be large.}
A crucial distinction with the previously considered cases is the presence of 
{\em continuous} chiral symmetries. At small 
$r(S_1)$, at weak coupling, the continuous chiral symmetries remain unbroken,
while  Abelian confinement of the Polyakov type
sets in, much in the same way as
in pure Yang--Mills or single-flavor theories. Linear confinement
coexists with the unbroken chiral symmetry in the quark
sector.\footnote{There is
no contradiction with the Casher argument \cite{casher} since the latter does not apply in
2+1 dimensions.}

In the strong coupling decompactification 
limit one expects non-Abelian confinement and spontaneous 
$\chi $SB. We study the dynamical origin and other
details of the $\chi $SB phenomenon within our theoretical framework. 
We observe a chiral phase transition in passing 
from small to large $r(S_1)$ in the multiflavor case. We conjecture that the scale of the 
$\chi $SB is tied with the  passage from Abelian to non-Abelian confinement, and is of the order  $L_{ \chi SB} \sim \Lambda^{-1}/N$. This surprising suppressed scale would be a natural 
scale of $\chi$SB were the center symmetry stable all the way down to arbitrarily small  $r(S_1)$.

\section{Theoretical framework}

The general design is as follows.
We consider SU$(N)$ Yang--Mills theories with $N_f$ flavors where $N_f>1$.
Each flavor is described by the Dirac fermion field in the complex representation
\beq
{\mathcal R} =\{ \rm F,\,\, S,\,\, AS,\,\, BF\}
\eeq
where F stands for fundamental, AS/S/BF stand for two index antisymmetric,
symmetric and bifundamental representations. We assume $N_f$ to be sufficiently small
so that asymptotic freedom is preserved and the theory at hand is below the lower boundary of the conformal window.  For simplicity we will focus on 
$N_f=2$.
The action for multiflavor QCD-like theories on $R_3\times S_1$ takes the form
\beq
S = \int_{R_3\times S_1}\, \frac{1}{g^2}\left[\frac{1}{2} \Tr F_{MN}^2 +
i\bar\Psi^a \slsh{D}\Psi_a
\right]
\label{action}
\eeq
where $a$ is the flavor index and $\slsh{D} = \gamma_M(\partial_M +iA_M)$
is the covariant derivative acting in the representation $\mathcal R$.
For
QCD(BF) the gauge group is SU$(N)\times$SU$(N)$, and the gauge part of the action 
(\ref{action}) must be replaced by
\beq
F_{MN}^2\to F_{1,MN}^2 + F_{2,MN}^2\,. 
\eeq

\begin{figure} 
\epsfxsize=4cm
\centerline{\epsfbox{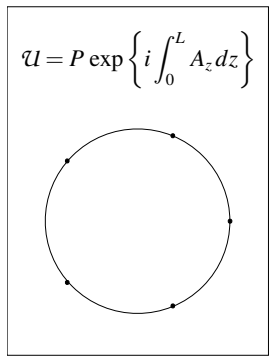}}
\caption{ The center-symmetric configuration of the eigenvalues of the Polyakov line for SU$(5)$.} \label{ndsv} 
\end{figure}

On a small cylinder  $r(S_1)\ll \Lambda^{-1}$,  one can deform the original  theory  by adding a  double-trace operator $P[U({\bf x}) ]$
where 
 \begin{equation}
P[U({\bf x}) ]= \frac{2}{\pi^2 L^4} \, \sum_{n=1}^{\left[\frac{N}{2}\right]}  d_n        
 \left|
\, {\rm Tr}\, U^n ({\bf x} )\right|^2 \,,
\label{fourma}
\end{equation}
 $d_n$ are numerical parameters of order one, and ${\left[...\right]} $ denotes the integer part of 
the argument in the brackets. 
The deformed action is 
\begin{equation}
S^{*} = S + \int_{R_3 \times S_1} P[U({\bf x}) ]\,.
\end{equation} 
For judiciously chosen  $d_n$, the center  symmetry remains unbroken in the vacuum
while -- due to weak  gauge coupling  {\it  and} center symmetric holonomy -- the gauge symmetry 
SU$(N)$ spontaneously breaks,
\beq
{\rm SU}(N)\to {\rm U}(1)^{N-1}\,.
\label{higgsed}
\eeq
The eigenvalues of the Polyakov line $\mathcal{U}$ in the vacuum 
have a regular pattern 
\beq
u_k = e^{\frac{2\pi i k}{N}}\,,\quad k = 0,1, ..., N-1
\eeq
 depicted in Fig.~\ref{ndsv}. 
$N-1$ diagonal gauge bosons -- photons -- remain perturbatively massless, while off-diagonal 
gauge bosons acquire masses $\sim 1/L$.
For  what follows it will be convenient to introduce
\beqn
\langle A_z \rangle 
&=&
 \frac{1}{L}{\rm diag}\left\{-\frac{2\pi [N/2]}{N},\,\,  -\frac{2\pi ([N/2]-1)}{N}, ....,\,
\frac{2\pi [N/2]}{N}\right\} ,
\label{12}
\eeqn
a matrix whose main diagonal is proportional to $\ln\,u_k$. If $N$ is odd, one of the eigenvalues is zero, and there is a fermionic mode which is massless. 

On fermions we will impose a  boundary condition with a U$(1)$-twist
\beq
\Psi^a (\mbox{\boldmath$x$}, x_4+L) = e^{2\pi i \omega}\, \Psi^a (\mbox{\boldmath$x$}, x_4)\,,
\qquad g^2_{4D} \ll \omega\ll 1\,.
\eeq
which is equivalent to turning on an overall U(1) Wilson line for the background holonomy.  
The U(1)-shifted holonomy  
 $\langle A_z \rangle - \omega \frac{2 \pi}{L} $ generates 
three-dimensional  real mass terms for  the fermion fields which  does  not break any of the chiral symmetries  (\ref{chiral}) inherent to the multiflavor theories on $R_4$.  This is unlike 
the complex four-dimensional  mass which would explicitly violate $\chi$S. The U$(1)$-twist
 $\omega = 0^+$ plays the role of an infrared regulator in loops with (otherwise massless)
fermions. 
 
 The chiral   symmetry group  of the action (\ref{action}) is   
\beq
\label{chiral}
\frac{
{\rm SU}(N_f )_L \times {\rm SU}(N_f )_R \times {\rm U}(1)_V \times Z_{2hN_f}
}{
Z_{N_f} \times Z_{N_f} \times Z_2} ~.
\eeq
The factors in the denominator eliminate double-counting of
the symmetries.  $2N_fh$ is the number of fermion zero modes in the background
of the Belavin--Polyakov--Schwarz--Tyupkin 
(BPST) instanton
\cite{BPST} on $R_4$. For the fermionic representations of interest
\beq
2h = \{2, \,\, 2N + 4, \,\, 2N - 4, \,\, 2N\} \quad {\rm for}\quad {\mathcal R} = \{{\rm F, S, AS, BF}\} .
\eeq
The action of  (\ref{chiral}) on the Weyl  fermions  is 
\beqn
&& {\rm SU}(N_f )_L : \qquad  \lambda^a\to (U\lambda)^a , \qquad
\bar\psi^a\to \bar\psi^a\,,
\nonumber\\[1mm]
&& {\rm SU}(N_f )_R : \qquad \lambda^a\to \lambda^a, \qquad \bar\psi^a\to (V\bar\psi )^a\,,
\nonumber\\[1mm]
&& {\rm U}(1)_V : \qquad  \lambda^a\to e^{i\delta} \lambda^a ,\qquad \bar\psi^a\to e^{i\delta}
\bar\psi^a\,,
\nonumber\\[1mm]
&& Z_{2hN_f} :  \qquad \lambda^a\to e^{\frac{2\pi i k}{2hN_f}} \lambda^a 
,\qquad \bar\psi^a\to e^{-\frac{2\pi i k}{2hN_f}}
\bar\psi^a\,.
\eeqn
Note that the four-dimensional Weyl fermion reduces to the Dirac fermion in the long-distance 
three-dimensional gauge theory. 
Thus, the relation between the four-dimensional Dirac fermion  and the three-dimensional Dirac fermion
obtained  upon reduction is 
\beq
\Psi = \left(
\begin{array}{c}
\lambda\\
\bar\psi
\end{array}
\right).
\eeq

At small $r(S_1)$ the eigenvalues of the Polyakov line
weakly fluctuate near their vacuum values depicted in Fig.~\ref{ndsv}. It is only the sum of the eigenvalues
that vanishes in the vacuum. At large $r(S_1)$ each eigenvalue is expected to wildly fluctuate
and average to zero. The same universality class hypothesis (say, for pure Yang--Mills)
is that the passage from one regime to another is smooth. 
In principle, the smoothness, as opposed to a phase transition 
on the way from small to large values of $r(S_1)$, can be tested on lattices.

\section{QCD$^*$ with two flavors}

For definiteness, let us start from the  case of fundamental fermions.
Relevant introductory material and notation can be found in \cite{SU2}.
There are $N-1$ distinct U(1)'s in this model, corresponding to $N-1$ distinct electric charges. Each component $\Psi_i^a$ ($i=1, ... , N$ and $a=1,2$) will be characterized by a set
of $N-1$ charges, which we will denote by $\mbox{\boldmath $q$}_{\Psi_i}$,
\beq
\mbox{\boldmath $q$}_{\Psi_i} = g\, \mbox{\boldmath $H$}_{ii} \equiv g\left( [H^1]_{ii}, [H^2]_{ii} \ldots, [H^{N-1}]_{ii} \right)
\,,\quad
i=1, ... , N\,,
\eeq
where $\mbox{\boldmath $H$}$ is the set of $N-1$ Cartan generators.
If $N$ is odd, $2N_f= 4$ fermion components remain massless (two $\psi^a$'s and two
$\lambda^a$'s). More exactly, these modes  are nearly massless, with mass $\omega/L$. 
Other components become massive and can be integrated out. 
If $N$ is even, to keep  fermions in the low-energy limit, 
we will have to tune  $2 \pi \omega = \pi/ N + 0^{+} $. 
The fermions that survive in the low-energy limit are charged and therefore appear in loops. 

The infrared dynamics can be described as  compact QED$_3$ with light fermionic matter.  
Due to gauge symmetry breaking (\ref{higgsed}) via a {\it compact  scalar} (the holonomy), 
there are $N$ types of elementary  instanton-monopoles. These topological
excitations are 
uniquely labeled by their magnetic charges valued in the affine root system  
$ \Delta^{0}_{\rm aff} = 
\{ \mbox{\boldmath $\alpha$}_1,  \mbox{\boldmath $\alpha$}_2, \ldots ,  \mbox{\boldmath $\alpha$}_N \}$ . 
The operators corresponding to the topological excitations are expressed in terms of 
the  dual variables for 
the photons  $ d \mbox{\boldmath $\sigma$} = *d \mbox{\boldmath $F$} $ by using Abelian duality. 
Summing over    the instanton-monopole contributions, 
the non-perturbative low-energy effective Lagrangian takes the form
\begin{eqnarray}
&& S^{\rm {QCD(F)}^*} =     \int_{R_3} \;   \Big[\,  
\,
 \frac{1}{4 g_3^2} \mbox{\boldmath $F$}^2  +   \frac{1}{g_{3}^2}
i \bar \Psi^a \left[ \gamma^{\mu}( \partial_{\mu} +  i \mbox{\boldmath $q$}_\Psi \mbox{\boldmath $A$}_{\mu} ) +\gamma^4\frac{2\pi i \omega}{L}\right] \Psi_a
\nonumber\\[3mm]
&& +  e^{-S_0} \, \Big( \tilde{\mu}\,   e^{i \, \mbox{\boldmath $\alpha$}_1 \mbox{\boldmath $\sigma$}} \,{\rm det}_{a,b=1,2}\{ \lambda^a \psi^b  \}+ 
     \mu\,  \sum_{\mbox{\boldmath $\alpha$}_j\in (\Delta^{0}_{\rm aff} - \mbox{\boldmath $\alpha$}_1 )}
             e^{i \mbox{\boldmath $\alpha$}_j\mbox{\boldmath $\sigma$} } 
             + {\rm H.c.}
  \Big)  + ...     \Big]  \,,
      \label{Eq:dQCD(F)}
\end{eqnarray}
where 
 $\mu $   and $\tilde\mu$ are 
dimensionful coefficients of the monopole operators and  the 
ellipses stand for higher order terms in
the topological $e^{-S_0}$ expansion and ignored massive modes.
The $N-1$  linearly independent instanton-monopole operators render all
$N -1$ dual photons  {\boldmath $\sigma$} massive, with masses proportional to $e^{-S_0/2}$.
This switches on the Polyakov linear confinement.\footnote{There are $N-1$ distinct strings.
In principle, they can break due to the fermion pair creation, but the breaking is exponentially suppressed.}

Let us now discuss the fermion sector of the low-energy theory
(\ref{Eq:dQCD(F)}). 
To establish the vacuum structure we note that
at distances larger than $e^{S_0/2}$, the photons are gapped and are pinned at the bottom of the instanton-monopole induced potential. Consequently, 
to find the symmetry of the vacuum we explore
the long-distance Lagrangian 
\begin{eqnarray}
 S^{\rm {NJL}} =   \!  \int_{R_3} \!   \left\{\,  \,
   \frac{1}{g_{3}^2}
i \bar \Psi^a \left[ \gamma^{\mu} \partial_{\mu} +
    \gamma^4\frac{2\pi i \omega}{L}\right] \Psi_a
+  e^{-S_0} \, \Big( \tilde{\mu}\,  
{\rm det}_{a,b=1,2}\{ \lambda^a \psi^b  \}+ 
            {\rm H.c.}
  \Big)   \right\} \,.
      \label{Eq:dQCD(F)2}
\end{eqnarray}
 In essence, it describes a
three-dimensional Nambu--Jona-Lasinio (NJL)
 model with ${\rm SU}(N_f )_L \times {\rm SU}(N_f )_R \times {\rm U}(1)_V$ chiral symmetry
 at the Lagrangian level. It is known that at arbitrarily weak coupling (in the
units of the cut-off scale), the chiral symmetry of the NJL model remains
unbroken. This is the phase of confinement without  $\chi$SB,
with  massless (or light)  fermions in the spectrum whose masslessness is protected by unbroken  $\chi$S.

As the coupling increases and approaches unity in the domain $r(S_1)\Lambda \sim 1$,
  the chiral symmetry (\ref{chiral}) is expected to spontaneously  break
down to the diagonal vector subgroup, SU$(N_f)_D$. This breaking must result in
massless $N_f^2 -1$ Nambu--Goldstone (NG) bosons. The $\chi$SB
phase  transition occurs at the boundary of the region of validity of the  low-energy theory
(\ref{Eq:dQCD(F)}).

Now let us extend the above discussion to fermions in the two-index representations
(still keeping $N_f=2$).

{\bf QCD(AS/S/BF)*:} In  QCD* theories with two-index fer\-mions in
the representations ${\cal R}=\{{\rm AS/S/BF}\}$  we also observe 
two different phases as is the case with  QCD(F)*. These are 
\begin{eqnarray}
\label{phases}
&&L < L_{ \chi SB} , \qquad {\rm with \; massless \; (or\;  light) \;  fermions},
 \nonumber\\[1mm]
&& L \geq L_{ \chi SB} , \qquad {\rm with \; massless  \; \; NG}\mbox{-}{\rm bosons} .
\end{eqnarray}
In the latter phase we have 
confinement with (continuous) $\chi$SB while in the former
confinement without  $\chi$SB. At any radius, the discrete chiral symmetry pattern is 
$Z_{2h} \rightarrow Z_2$ \cite{SU2}, probed  by a determinantal, continuos   $\chi$S-singlet order parameter, 
$\langle \det \lambda^a \psi^b \rangle$. Thus, at small $S_1$, these  theories  have $h$ isolated vacua and at large $r(S_1)$, $h$ isolated coset spaces. 
Below,  we  highlight  the differences in the analyses of two- and one-index 
representation fermions.  We take  QCD(BF)* as our main example due to its simplicity. 
QCD(AS/S) analysis is analogous, up to minor differences that can be filled in by using  Ref. \cite{SU2}.  

In two-flavor QCD(BF)*, the $4N$ zero modes of the BPST instanton (which can be viewed as $N$
instanton-monopoles) split into $N$ groups of four zero modes each \cite{Poppitz:2008hr}.  (This is the reason why the instanton-monopoles must play a more prominent role on $S_1 \times R_3$ than 
the four-dimensional  BPST instanton.)
 Thus, unlike  QCD(F)*, the instanton-monopoles  appearing at the order 
$e^{-S_0}$ do not cause confinement, but may induce $\chi$SB.  
The magnetic bions which appear at the order  
  $e^{-2S_0}$ lead to confinement, and mass gap for the gauge field fluctuations. The  leading  
  monopole and bion induced  nonperturbative effects are 
   \begin{eqnarray}
  L^{\rm QCD(BF)^*}_{\rm nonpert} &= 
&    \sum_{\alpha_{i} \in \Delta_{\rm aff}^{0}}\Big[  \tilde {\mu} e^{ -S_{0}} 
     ( e^{+i \mbox{\boldmath $\alpha$}_i\,  \mbox{\boldmath $\sigma$}_1 }  + 
     e^{+i \mbox{\boldmath $\alpha$}_i\,  \mbox{\boldmath $\sigma$}_2 }) 
     ({ \det_{a, b} \lambda_i^a  \psi_i^b +
 \det_{a,b} \lambda_{i+1}^a  \psi_{i+1}^b   } )\, 
             \nonumber\\[2mm]    
 &+&  \mu  e^{-2S_0}    
 \Big[      c_1 \Big( 
     e^{i ( \mbox{\boldmath $\alpha$}_i \, - \mbox{\boldmath $\alpha$}_{i-1} \,) \mbox{\boldmath $\sigma$}_1} + e^{i ( \mbox{\boldmath $\alpha$}_i
     \, - \mbox{\boldmath $\alpha$}_{i-1} \,) \mbox{\boldmath $\sigma$}_2}
     \Big)      
     \nonumber\\[2mm]     
& + &    c_2 \Big( 
2 e^{i ( \mbox{\boldmath $\alpha$}_i\, \mbox{\boldmath $\sigma$}_1 -
 \mbox{\boldmath $\alpha$}_i\,  \mbox{\boldmath $\sigma$}_2) } + 
  e^{i ( \mbox{\boldmath $\alpha$}_i\, \mbox{\boldmath $\sigma$}_1 -
  \mbox{\boldmath $\alpha$}_{i-1}\,  \mbox{\boldmath $\sigma$}_2) }
  + e^{i ( \mbox{\boldmath $\alpha$}_i \, \mbox{\boldmath $\sigma$}_1 -
\mbox{\boldmath $\alpha$}_{i+1}\,   \mbox{\boldmath $\sigma$}_2) }   
  \Big)             
       \Big]          
       +  {\rm  H.c.}
         \nonumber\\
           \end{eqnarray}
Consequently, the   gauge structure of the theory undergoes a two-stage breaking:\footnote{In the second stage of the gauge structure reduction,  the gauge group is not Higgsed, but, rather, the dual photons acquire   gauge invariant masses.}
  \begin{eqnarray}
{\rm SU}(N) \times {\rm SU}(N)  \; \; \; \stackrel{\rm  Higgsing}{\longrightarrow} \; \; \; 
 [{\rm U}(1)]^{N-1} \times  [{\rm U}(1)]^{N-1} 
 \; \; \; \stackrel{\rm nonperturbative}{\longrightarrow}  Z_N  \,.
\label{pattern1}
  \end{eqnarray} 
  The discrete gauge group  (DGG) appearing at the final stage 
  is another important difference between various QCD$({\cal R})^*$ theories. 
In  general, the    gauge symmetry breaking pattern pertinent to QCD$({\cal R})^*$ theories
can be described as
\begin{equation}
\label{chain}
G  \; \; \; \stackrel{\rm  Higgsing}{\longrightarrow} \; \; \;  {\bf Ab}(G)   
 \; \; \; \stackrel{\rm nonperturbative}{\longrightarrow}  
{\rm DGG}({\cal R})
\end{equation}
 where   ${\bf Ab}(G)$ is the Abelian gauge structure, as in the Seiberg--Witten  theory, 
 and  DGG  is the discrete gauge group which survives in the infrared. 
  ${\rm DGG}$ is equal to  $Z_{\kappa}$  where   ${\kappa}$ is determined by the representation of   massless fermions. For even $N$ 
 \beq
 \kappa= \{N, N, 2 , 2, 1\}, \;\; {\rm for}\;\;  {\cal R}= \{\rm Adj, BF, AS, S, F\},
 \eeq 
while for odd $N$,   $\kappa= \{N, N, 1, 1, 1\}$, respectively. Note that DGG is a 
subgroup of the center group and can be obtained as  the  quotient of  the center group by equivalences imposed by massless matter.   In  QCD$({\cal R})^*$, 
 the charges which are non-neutral under  DGG=$ Z_{\kappa}$ are confined.  
  This leads to  the second difference between QCD(F)* and   QCD$({\cal R})^*$ where 
  ${\cal R}= \{\rm BF, AS,S, Adj\}$. In the first problem, strictly speking,  the strings can break.
In the latter case,  the  area law behavior of large Wilson loops 
(typically due to magnetic bions) is exact. 
  This guarantees that charges nonvanishing under DGG  
  are confined. Nonetheless, the gauge fluctuations are gapped in both  cases. 

 As stated above  (see (\ref{phases})), the QCD$({\cal R})^*$ theories possess two phases of confinement,  with and without $\chi$SB. Returning to QCD(BF)*, at distances larger than 
 $\sim e^{S_0}$, the photons are gapped, and the vacuum structure is determined by the  fermion
 action 
 \begin{eqnarray}
&& S= \int_{R^3}  \sum_{i=1}^{N}  \left[
    i \overline \Psi^a_i  \left[ \gamma^{\mu} \partial_{\mu} +
    \gamma^4\frac{2\pi i \omega}{L}\right]
     \Psi_{a,i} 
      +  2 
      \tilde{\mu} e^{-S_0}  \Big( \det_{a, b} \lambda_{i}^a \psi_{i}^b  + { \rm h.c } \Big)   \right] \; .
\label{verylong}
\end{eqnarray}
This is again, an NJL-type model, with the same consequences 
as those discussed  around (\ref{phases}). For $L < L_{\chi SB}$, we have massless or light fermions in the spectrum -- no complex Dirac mass is generated. For $L > L_{\chi SB}$,  the chiral symmetry 
is broken via the bilinear $ \langle \lambda^a \psi^b \rangle \sim \delta^{ab}$, inducing a 
{\it complex} four-dimensional Dirac mass for the fermion.  This phase possesses massless 
NG-bosons due to $\chi$SB. 

\section{Abelian to non-Abelian confinement}

In the small-$r(S_1)$ regime, the mechanism of confinement is Abelian,  by virtue of   $\bf Ab (G)$ 
in the gauge structure chain   (\ref{chain}).\footnote{In fact, this is true for all the analytically controlled mechanisms of confinement known  so far, including the Seiberg--Witten theory \cite{SW1} or Polyakov's mechanism \cite{Pol}.  The reason for the applicability of the semiclassical analysis is the appearance of 
an ${\bf Ab} (G)$ structure at some  length scale.} At  $e^{-S_0} \sim r(S_1)\Lambda \sim 1$, 
 one looses  the separation of scales between the lightest $W$-bosons and the heaviest 
 nonperturbatively gapped photons, so that the long distance theory based on  $\bf Ab (G)$
 looses its validity.  This is also the scale at which one expects the long-distance NJL-model
to induce the $\chi$SB.   We believe that, the scale of the 
passage from the Abelian to non-Abelian confinement and that of the chiral phase transition
 are parametrically tied up, and the bilinear chiral order parameter probes both. 

This suggests that, in multiflavor QCD-like gauge theories, 
 confinement without $\chi$SB is a property of the Abelian confinement, whereas, 
 continuous  $\chi$SB is associated with non-Abelian confinement. 
 
 One other surprising aspect of this  chiral transition is its scale.  In this work we dealt with small $N$, small $N_f$ theories. However, if we let $N$ to be arbitrarily large, 
 we would observe that the scale of the chiral transition is a {\it sliding } (or  suppressed) scale as a function of $N$.  We found, by either an order of magnitude estimate based on  the NJL Lagrangian,
   or  by employing more powerful large-$N$ volume independence theorem of 
   the center symmetric theories,  that    $L_{ \chi SB}  \sim {\Lambda^{-1}}/{N}$.   
   This also means that in the $N=\infty$ limit,  the region of Abelian confinement shrinks to zero, in compliance with the volume independence. 
(See section 5 of Ref.~\cite{SU3} and references therein.) The emergence of such $N$-suppressed  physical  scales in QCD-like theories  is  rather surprising by itself, and is outside the reach of   perturbation theory and non-perturbative 
holographic (supergravity) constructions.  It is  testable by numerical lattice simulations.

\section*{Acknowledgments}

We are grateful to E. Poppitz for useful discussions. M.S. is
supported in part
by DOE Grant DE-FG02-94ER-40823.
The work of  M.\"U. is supported by the U.S.\ Department of Energy Grant DE-AC02-76SF00515.

\hspace{1cm}

\end{document}